\documentclass[%
reprint,
superscriptaddress,
amsmath,amssymb,
aps,
]{revtex4-2}

\usepackage{graphicx}% Include figure files
\usepackage{dcolumn}% Align table columns on decimal point
\usepackage{bm}% bold math
\usepackage{mathrsfs}
\usepackage{float}
\usepackage{physics}
\usepackage{siunitx}
\usepackage[colorlinks,linkcolor=blue,anchorcolor=blue,urlcolor=blue,citecolor=blue]{hyperref}
\begin{document}
	%\linenumbers % 开启行号

\preprint{APS/123-QED}

\title{Quantum-Optical Bound States in the Continuum}
\author{Ruo Kun Cai}
\affiliation{College of Science, National University of Defense Technology, Changsha, Hunan 410073, China\\}

\author{Zhi Jiao Deng}
\email{dengzhijiao926@hotmail.com}
\affiliation{College of Science, National University of Defense Technology, Changsha, Hunan 410073, China\\}
\affiliation{Hunan Key Laboratory of Mechanism and Technology of Quantum Information, Changsha, Hunan 410073, China\\} 

\author{Chun Wang Wu}
\email{cwwu@nudt.edu.cn}
\affiliation{College of Science, National University of Defense Technology, Changsha, Hunan 410073, China\\}
\affiliation{Hunan Key Laboratory of Mechanism and Technology of Quantum Information, Changsha, Hunan 410073, China\\}

\author{Ping Xing Chen}
\affiliation{College of Science, National University of Defense Technology, Changsha, Hunan 410073, China\\}
\affiliation{Hunan Key Laboratory of Mechanism and Technology of Quantum Information, Changsha, Hunan 410073, China\\}
\affiliation{Hefei National Laboratory, Hefei, Anhui 230088, China\\}

\date{\today}% It is always \today, today,
             %  but any date may be explicitly specified

\begin{abstract}
Bound states in the continuum (BICs) are counterintuitive localized states that lie within the continuum of extended states. While extensively realized and utilized in classical wave systems, it is still unclear what a close analog of BICs would be, and how to extract their experimental signature in quantum-optical settings—where the wave field itself is quantized into bosonic excitations. Here, we present a paradigmatic quantum-optical model consisting of a driven multi-level Jaynes-Cummings (JC) system, featuring few quantum degrees of freedom yet capable of hosting a BIC. Using the concept of a Fock-state lattice (FSL), this model can be mapped to an extended structure comprising two semi-infinite inhomogeneous Su-Schrieffer-Heeger (SSH) chains coupled to a common continuum. An appropriate quantum superposition of two topological zero modes from the separate chains forms a BIC that remains perfectly localized in the Fock-state dimension within the continuum spectrum, due to complete decoupling from the common continuum via destructive quantum interference. We further develop a method to extract the spectroscopic signature of the BIC—a discrete peak embedded in a continuous background—by Fourier-transforming the time-dependent dynamics of the system’s chiral-symmetry operator. A highly feasible experimental proposal using a single trapped ion is provided. Our work bridges BIC physics with quantum optics, opening a pathway to harnessing such exotic states at the quantum limit.

\end{abstract}

\maketitle

\section{introduction}Bound states in the continuum (BICs) are physical states that, despite lying within the continuous spectrum of extended or radiative states, remain perfectly spatially confined—a property that defies conventional wisdom \cite{zs1-2016,zs2-2021,zs3-2021,zs4-2024,zs5-2026}. Although originally proposed in the early days of quantum mechanics by von Neumann and Wigner in 1929 \cite{von-1929}, BICs are now recognized as a universal wave phenomenon, observed in diverse classical systems including optical waves \cite{zs2-2021,zs4-2024,zs5-2026}, acoustic waves \cite{acoustic1,acoustic2,acoustic3,acoustic4-prl,acoustic5-prl}, water waves \cite{water1-1996,water2-2009,water3-2011} and elastic waves in solids \cite{elastic1-1976,elastic2-1992}. Among these, optical BICs have garnered the greatest interest and have been experimentally explored in a variety of classical platforms, from photonic crystals \cite{optical1-2008,optical2-2012,optical3-2013,optical4-2014,optical5-2022,optical6-2022,optical7-2025} and optical waveguides \cite{waveguide1-2011,waveguide2-2013,waveguide3-2009,waveguide4-2013,waveguide5-2020,waveguide6-2021} to metasurfaces \cite{metasurfaces1-2018,metasurfaces2-2018,metasurfaces3-2019,metasurfaces4-2019,metasurfaces5-2019,metasurfaces6-2024} and exciton–polariton systems \cite{eps1-2018,eps2-2024,eps3-2025,optical7-2025}. These efforts have not only advanced the understanding of wave–matter interactions but have also spurred breakthrough applications, e.\,g., ultrahigh‑Q lasers \cite{lasers1-2017,lasers2-2017,lasers3-2020,eps3-2025,optical7-2025}, ultra‑sensitive sensors \cite{metasurfaces1-2018,sensors1-2017,metasurfaces5-2019,metasurfaces4-2019,sensors4-2025}, and enhanced high-harmonic generation \cite{highhamor1-2018,metasurfaces3-2019,highhamor2-2019,highhamor3-2020,highhamor4-2020}.

Despite the extensive study and application of classical BICs, the development of their quantum counterparts has long lagged behind \cite{zs1-2016}. To date, phenomena related to quantum BICs have been observed in only a handful of ultracold atomic-molecular experiments. Examples include the suppression of autoionization linewidth in Rydberg atoms \cite{Rydberg1-1985,Rydberg2-1985,Rydberg3-2002}, cold-atom collisions mediated by Feshbach resonances \cite{coldatom1-2014,coldatom2-2026}, and corner states of the atomic-momentum lattice \cite{momenlat1-2025}. Recent theoretical studies have also suggested the potential existence of quantum BICs in paradigmatic condensed matter models, such as quantum Hall systems \cite{quhall-2013}, Hubbard models \cite{hubbard1-2012,hubbard2-2013}, temporally modulated materials \cite{modul1-2014,modul2-2013,modul3-2020}, and strongly correlated many-body systems \cite{manybody1-2024}. Nevertheless, a striking lack of integration persists between current research on quantum BICs and the field of quantum optics—a field endowed with sophisticated experimental techniques that offer rich potential. The realization and detection schemes for a genuine quantum-optical BIC remain entirely unexplored.

To bridge this gap, we investigate feasible routes towards realizing BICs within a quantum-optical setting. A common prerequisite for BICs is an extended structure, which poses a particular challenge for typical quantum-optical systems with few physical degrees of freedom. The recently developed concept of Fock-state lattices (FSLs) \cite{fsl1-2022,fsl2-2024,fsl3-2023,fsl4-2024,fsl5-2025,fsl6-2023,fsl7-2010,fsl8-2016,fsl9-2011,fsl10-2016,fsl11-2021,fsl12-2023,fsl13-2024,fsl14-2025} provides a pathway forward, mapping the infinite-dimensional Fock space of a bosonic mode onto the sites of an extended lattice and thereby engineering the necessary spatial extension. Building on this idea, we start from a driven multi-level Jaynes-Cummings (JC) model and, via a Fock-state mapping, construct a composite architecture comprising two Su-Schrieffer-Heeger (SSH) chains simultaneously coupled to a common continuum. A key feature of this structure is the exact decomposition of its Hamiltonian, which allows an appropriate superposition of the two topological zero modes from the two SSH chains to completely decouple from the continuum. This results in a quantum BIC, embedded within the continuum. To characterize and detect this state, we develop a spectroscopic protocol based on the dynamics of the system’s chiral operator, which clearly reveals the BIC as a discrete spectral peak embedded in a continuous background. Moreover, we present a highly feasible experimental scheme using a single trapped ion. This work not only conceptually merges BIC physics with quantum optics but also provides a concrete feasible route for preparing and characterizing BICs on a fully controllable quantum platform, laying the foundation for their future potential applications in quantum information and quantum-enhanced metrology.

\begin{figure}[t]
	\includegraphics[width=0.48\textwidth]{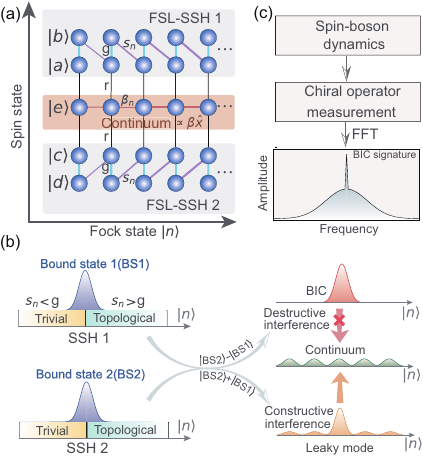}
	\centering
	\caption{\label{fig:BIC_FSL_SSH}Schematic of the spin-boson lattice model and the formation of a quantum-optical BIC. (a) Mapping of the driven JC model onto a FSL, comprising two semi-infinite anisotropic SSH chains coupled to a central chain with a continuous spectrum. (b) Symmetric coupling of the localized bound states (in each SSH chain) to the common continuum. Destructive and constructive quantum interference between these channels gives rise to a BIC and a leaky mode, respectively. (c) Spectroscopic detection protocol: Fourier analysis of the chiral symmetry operator’s time evolution reveals the BIC signature—a discrete spectral peak within a continuous background.}
\end{figure}

\section{basic model and mechanism for quantum-optical BIC}
The model we consider is a generalized JC-type quantum-optical system, consisting of a high-dimensional pseudo-spin with five levels $\{|i\rangle, i = a, b, c, d, e\}$ and a bosonic mode. The system incorporates resonant drives applied to both the spin and the bosonic mode, as well as JC-type couplings between them. Its Hamiltonian is written as: 
\begin{equation}
	\begin{aligned}
		\hat{H} &= g \bigl(\hat{\sigma}_x^{a,b}+\hat{\sigma}_x^{c,d}\bigr)+r\bigl(\hat{\sigma}_x^{a,e}+\hat{\sigma}_x^{c,e}\bigr)\\
		&+s\bigl(\hat{\sigma}_-^{a,b}\hat{a}^\dagger+\hat{\sigma}_-^{c,d}\hat{a}^\dagger+\rm{H.c.}\bigr)\\
		&+\sqrt{2}\beta |e\rangle\langle e| \hat{x},
	\end{aligned}
	\label{Hamiltonian}
\end{equation}
where $\hat{\sigma}_x^{i,j}=|i\rangle\langle j|+\rm{H.c.}$ are the spin Pauli-x operators for the transition between states $|i\rangle$ and $|j\rangle$, $\hat{\sigma}_-^{i,j}=|i\rangle\langle j|$ are the corresponding spin lowering operators, $\hat{x} = \frac{1}{\sqrt{2}}(\hat{a}+\hat{a}^\dagger)$ is the bosonic coordinate operator, and $\hat{a}$ and $\hat{a}^\dagger$ are the bosonic annihilation and creation operators in the Fock-state basis $\{|n\rangle\}$, respectively. In the Hamiltonian, the first two terms describe the spin drives with strengths $g$ and $r$, the third term represents the JC-type couplings between the spin and the bosonic mode with strength $s$, and the fourth term corresponds to a state-dependent bosonic drive with strength $\sqrt{2}\beta$, which takes effect only when the spin is in state $|e\rangle$. This model is possible to be implemented on various quantum-optical platforms, e.\,g., trapped ions \cite{trapion1-2015} and cavity quantum electrodynamics \cite{edynamics1-2002}.

Inspired by the concept of FSL \cite{fsl3-2023}, where the bosonic Fock-state space and the spin-state space are reinterpreted as synthetic spatial dimensions, the driven JC model described by Eq.\,(\ref{Hamiltonian}) can be mapped onto a two-dimensional lattice structure. The lattice sites are represented by the basis states $\{|i\rangle|n\rangle, i = a, b, c, d, e; n=0,1,2,\ldots,\infty\}$ [Fig.\,\ref{fig:BIC_FSL_SSH}(a)]. In this mapping, the spin-drive terms in $\hat{H}$ provide couplings along the vertical direction of the lattice, the JC-type couplings introduce diagonal hoppings, and the state-dependent bosonic drive induces couplings along the horizontal direction between sites associated with the $|e\rangle$ state. It is important to note that the diagonal and horizontal couplings are site-dependent, with strengths $s_n=s\sqrt{n+1}$ and $\beta_n=\beta\sqrt{2(n+1)}$, respectively---a direct consequence of the properties of the bosonic annihilation and creation operators. One can observe that the subsets of sites $\{|a\rangle|n\rangle, |b\rangle|n\rangle, n=0,1,2,\ldots,\infty\}$ and $\{|c\rangle|n\rangle, |d\rangle|n\rangle, n=0,1,2,\ldots,\infty\}$ form two identical semi-infinite, and anisotropic SSH chains, similar to the type analyzed in Ref.\, \cite{fsl14-2025}. Meanwhile, the subset $\{|e\rangle|n\rangle, n=0,1,2,\ldots,\infty\}$ constitutes a one-dimensional chain with a continuous spectrum, whose Hamiltonian corresponds to that of a driven quantum harmonic oscillator. These two SSH chains are connected to the central 1D chain via coupling channels of strength $r$, linking sites that share the same quantum number $n$.

The formation mechanism of the quantum-optical BIC is schematically illustrated in Fig.\,\ref{fig:BIC_FSL_SSH}(b). As analyzed in Ref.\,\cite{fsl14-2025}, a semi-infinite anisotropic SSH model implemented in a FSL with site-dependent coupling hosts a domain wall. This domain wall separates the chain into a topologically trivial region ($s_n<g$) and a topologically nontrivial region ($s_n>g$), giving rise to a robust localized bound state at the interface. In our constructed lattice, each of the two identical SSH chains supports such a localized bound state, denoted as $|\rm{BS1}\rangle$ and $|\rm{BS2}\rangle$, respectively. Both bound states are symmetrically coupled to the central one-dimensional chain (the continuum) with strength $r$. Quantum interference occurs between these two coupling channels. Consequently, the specific superposition $|\rm{BS1}\rangle-|\rm{BS2}\rangle$ completely decouples from the continuum due to destructive interference, thereby forming a perfectly localized state in the Fock-state dimension, corresponding to a BIC. The orthogonal superposition, $|\rm{BS1}\rangle+|\rm{BS2}\rangle$, experiences enhanced coupling to the continuum via constructive interference, evolving into a leaky mode that extends in the Fock-state dimension. To detect this BIC, we propose a spectroscopic protocol [Fig.\,\ref{fig:BIC_FSL_SSH}(c)]: prepare a suitable initial state of the driven JC model, measure the time evolution of the system's chiral symmetry operator, and then perform a Fourier analysis. This procedure yields the system's energy spectrum, in which the hallmark signature of the BIC—a discrete spectral peak embedded within a continuous background—can be clearly observed.

\section{exact decomposition of the Hamiltonian}
From the lattice configuration in Fig.\,\ref{fig:BIC_FSL_SSH}, it is evident that within a sector of fixed boson number $n$, the driven JC model possesses a specific exchange symmetry. This symmetry interchanges the spin states as $|a\rangle \leftrightarrow |c\rangle$ and $|b\rangle \leftrightarrow |d\rangle$, while leaving $|e\rangle$ unchanged. We define the corresponding symmetry operator $\hat{S}$ acting on the spin states as follows: $\hat{S}|a\rangle = |c\rangle$, $\hat{S}|c\rangle = |a\rangle$, $\hat{S}|b\rangle = |d\rangle$, $\hat{S}|d\rangle = |b\rangle$, and $\hat{S}|e\rangle = |e\rangle$. One can readily verify that $\hat{S}$ is both Hermitian and unitary, and that it commutes with the system Hamiltonian, $[\hat{S}, \hat{H}] = 0$. Since $\hat{S}^2 = I$, its eigenvalues are $\lambda = \pm 1$. Consequently, the entire Hilbert space can be decomposed into two uncoupled subspaces, each associated with a distinct eigenvalue of $\hat{S}$. Accordingly, the system Hamiltonian can be expressed as the direct sum of two independent Hamiltonians acting on these two orthogonal subspaces.

We now construct a set of basis states that diagonalize the symmetry operator $\hat{S}$. Specifically, we define:
\begin{equation}
	\begin{gathered}
		\ket{a',n} = \frac{1}{\sqrt{2}} \left( -\ket{a,n} + \ket{c,n} \right),\\
		\ket{b',n} = \frac{1}{\sqrt{2}} \left( -\ket{b,n} + \ket{d,n} \right),\\
		\ket{c',n} = \frac{1}{\sqrt{2}} \left(\ket{a,n} + \ket{c,n} \right),\\
		\ket{d',n} = \frac{1}{\sqrt{2}} \left(\ket{b,n} + \ket{d,n} \right),\\
		\ket{e',n} = \ket{e,n}.
	\end{gathered}
\end{equation}
In this new basis, $|a'\rangle$ and $|b'\rangle$ are eigenstates of $\hat{S}$ with eigenvalue $-1$, whereas $|c'\rangle$, $|d'\rangle$ and $|e\rangle$ correspond to eigenvalue $+1$. The corresponding projection operators onto the $\lambda = \pm 1$ subspaces are $\hat{P}_{\pm} = \frac{1}{2}\left(I \pm \hat{S}\right)$. Acting with these projectors on the full Hamiltonian leads to 
\begin{equation}
\hat{H} = \hat{P}_{+} \hat{H} \hat{P}_{+} + \hat{P}_{-} \hat{H} \hat{P}_{-} = \hat{H}_1 + \hat{H}_2,
\label{HamiltonianDecomposition}
\end{equation}
where
\begin{equation}
	\hat{H}_1 = g \hat{\sigma}_x^{a'b'} + s \bigr(\hat{a}^\dagger \hat{\sigma}_-^{a'b'}+\rm{H.c.}\bigr)
	\label{eq:H1} 
\end{equation}
and 
\begin{equation}
	\begin{aligned}
	\hat{H}_2 &= g \hat{\sigma}_x^{c'd'} + s \bigr(\hat{a}^\dagger \hat{\sigma}_-^{c'd'} +\rm{H.c.}\bigr) \\
	&+\sqrt{2}r \hat{\sigma}_x^{c'e}
	+\sqrt{2}\beta\ket{e}\bra{e}\hat{x}.
	\end{aligned}
	\label{eq:H2}
\end{equation}

\begin{figure}[t]
	\includegraphics[width=0.48\textwidth]{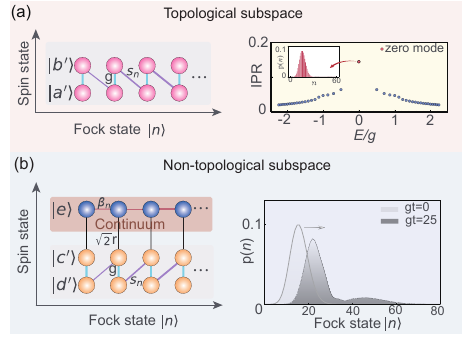}
	\centering
	\caption{\label{fig:subspace_localization}Exact decomposition of the Hilbert space and lattice structure. (a) Topological subspace: mapping onto a semi-infinite anisotropic SSH chain, which hosts a topological zero mode localized at the domain wall. The markedly higher inverse participation ratio (IPR) of this mode and its confined probability distribution in the Fock-state basis (inset) confirm its localized character. (b) Trivial (non-topological) subspace: the SSH chain coupled to a continuum. The original topological zero mode becomes unstable due to leakage into the continuum, manifested by the broadening and diffusion of the probability distribution $p(n)$ during time evolution. Parameters: $g$=1, $s$=0.25, $r$=1, and $\beta$=0.3.}
\end{figure}

By reinterpreting this new set of basis states as lattice sites, the full lattice structure shown in Fig.\,\ref{fig:BIC_FSL_SSH} can be decomposed into two decoupled sublattices, described by the Hamiltonians $\hat{H}_1$ and $\hat{H}_2$, respectively (see Fig.\,\ref{fig:subspace_localization}). The sublattice associated with $\hat{H}_1$ again maps onto a semi‑infinite, anisotropic FSL‑SSH chain, which hosts a robust topological bound state localized at the domain wall. In contrast, the sublattice corresponding to $\hat{H}_2$ describes an SSH chain coupled to a one‑dimensional chain with a continuous spectrum via a coupling of strength $\sqrt{2}r$; this coupling breaks the topological character of the original SSH chain. Consequently, the two subspaces obtained from the symmetry‑based decoupling correspond to a topologically nontrivial subspace and a topologically trivial subspace, respectively. 

As illustrated in Ref.\, \cite{fsl14-2025}, the eigenstates of $\hat{H}_1$, which governs the topological subspace, can be obtained analytically. Their explicit forms are given by:
\begin{align}
	\ket{\phi_0} &= \hat{D}(\gamma) \ket{a', 0} = \ket{a', \gamma}, \label{eq:zeromode}\\
	\ket{\phi_{\pm n}} &= \hat{D}(\gamma) \frac{\ket{b', n} \pm \ket{a', n+1}}{\sqrt{2}},\label{eq:eigenmode}
\end{align}
with the corresponding eigenvalues
\begin{equation}
	E_0=0, \quad E_{\pm n}=\pm s\sqrt{n+1},
	\label{eq:eigenenergy}
\end{equation}
where $\hat{D}(\gamma)$ is the bosonic displacement operator and $\gamma = -g/s$. The state $\ket{\phi_0}$ is the zero-energy mode, corresponding to a stable localized bound state at the domain wall. Its localized nature is manifested by a significantly larger inverse participation ratio (IPR) compared to other eigenmodes [Fig.\,\ref{fig:subspace_localization}(a)]. The IPR \cite{IPR} , commonly used to quantify the degree of localization of a quantum state, is defined as $\mathrm{IPR}=\sum_n p^2(n)$, where $p(n)$ is the probability at the sites with Fock-state number $n$. In the inset of Fig.\,\ref{fig:subspace_localization}(a), we also plot the probability distribution $p(n)$ of the zero mode, which provides a direct visual confirmation of its localized character. 

The Hamiltonian acting on the topologically trivial subspace describes a semi-infinite SSH chain coupled to a continuum. According to Fano resonance theory \cite{fano}, the discrete eigenstates of the SSH chain hybridize with the continuum, thereby acquiring finite lifetimes. In Fig.\,\ref{fig:subspace_localization}(b), the plotted probability distributions show that a state initially prepared in the zero mode of the SSH chain exhibits significant spreading at later times due to coupling with the continuum. This dynamical evolution directly demonstrates that the originally localized zero mode delocalizes in the Fock‑state dimension, degrading into a leaky mode.

Crucially, because $\hat{H}_1$ and $\hat{H}_2$ act on orthogonal subspaces, the localized zero mode in the topological subspace remains completely decoupled from the continuum of the trivial subspace. Consequently, this zero mode lies within the continuous spectrum provided by the trivial subspace yet exhibits no leakage into it, thus forming a perfect, non‑decaying quantum-optical BIC. Notably, this zero mode corresponds exactly to the superposition state $|\rm{BS1}\rangle-|\rm{BS2}\rangle$ shown in Fig.\,\ref{fig:BIC_FSL_SSH}, confirming directly the quantum‑interference mechanism responsible for the BIC formation.

\section{spectroscopic protocol for detecting the quantum-optical BIC}
From the analysis in the preceding section, preparing the system in the BIC state leads to no time evolution and no leakage into the continuum, which directly confirms the existence of the BIC from a dynamical perspective. However, such a dynamical verification only reflects the localized stability of the BIC. In this section, we aim to construct a detection protocol that can directly extract the spectral signature of the BIC—namely, a discrete spectral peak embedded within a continuous background. This signature offers a more intuitive correspondence to BICs in classical optical systems. In the following, we demonstrate that this characteristic signal can be obtained by measuring the dynamics of the system’s chiral-symmetry operator followed by a Fourier transform \cite{fft,fft2}. It should be noted that this protocol is effective only under the condition of weak coupling between the SSH chains and the continuum.

For the topological subspace, the chiral symmetry operator of $\hat{H}_1$ is 
\begin{equation}
	\hat{C}_1 = \left(|a'\rangle\langle a'| - |b'\rangle\langle b'|\right) \otimes I_{\text{boson}},
\end{equation}
satisfying $\hat{C}_1 \hat{H}_{1} \hat{C}_1 = -\hat{H}_{1}$, where $I_{\text{boson}}$ is the identity operator for the bosonic degree. This implies that for any eigenstate  $|\phi_\alpha\rangle$ with energy $E_\alpha$ in Eqs.\,(\ref{eq:zeromode})-(\ref{eq:eigenenergy}), $\hat{C}_1|\phi_\alpha\rangle = -|\phi_{-\alpha}\rangle$ is an eigenstate with energy $-E_\alpha$. If the system is initialized in $|\Psi_T(t=0)\rangle =|a',0\rangle$, which can be expanded in the eigenstates of $\hat{H}_1$ as
\begin{equation}
	|a',0\rangle = \sum_\alpha C_\alpha |\phi_\alpha\rangle,
\end{equation}
from Eqs.\,(\ref{eq:zeromode})-(\ref{eq:eigenenergy}), we have $C_{-\alpha}=-C_{\alpha}$ and $E_{-\alpha}=-E_{\alpha}$,the time-dependent expectation value of $\hat{C}_1$ then takes the form
\begin{equation}
	\begin{split}
		\langle \hat{C}_1(t) \rangle &=\sum_{\beta,\alpha} C_\beta^* C_\alpha \langle \phi_\beta | \hat{C}_1 | \phi_\alpha \rangle e^{-i(E_\alpha - E_\beta)t}\\
		&=\sum_\alpha |C_\alpha|^2 e^{-2i E_\alpha t}.
	\end{split}
\end{equation}

\begin{figure}[t]
	\includegraphics[width=0.48\textwidth]{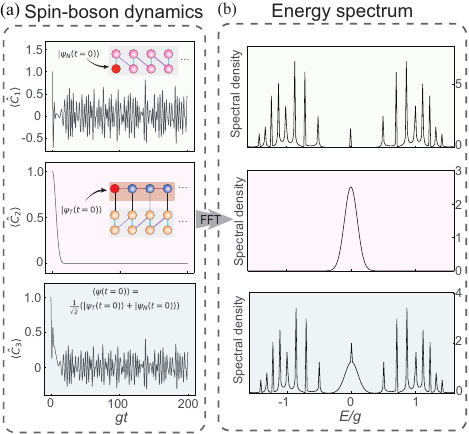}
	\centering
	\caption{\label{fig:dynamics_spectrum}Spectroscopic protocol for detecting the quantum-optical BIC. (a) Time evolution of the chiral-symmetry-operator expectation values for three different initial states: the topological zero mode $|a',0\rangle$ (top), $|e,0\rangle$ residing in the trivial subspace (middle), and their coherent superposition (bottom). (b) Corresponding spectra obtained by Fourier transform, showing a zero-mode discrete peak (top), a broad continuum of the trivial subspace (middle), and the definitive BIC signature—a sharp peak embedded in a continuous background (bottom). Parameters: $g$=1, $s$=0.5, $r$=0.01, and $\beta$=0.1.}
\end{figure}

Fourier transforming $\langle\hat{C}_1(t)\rangle$ gives the energy spectrum
\begin{equation}
	\begin{split}
F(\omega) &= \int_{-\infty}^\infty \langle\hat{C}_1(t)\rangle e^{-2i\omega t} dt\\
&= \pi \sum_{\alpha} |C_{\alpha}|^2 \delta(\omega + E_{\alpha}),
	\end{split}
\end{equation}
which exhibits peaks at $\omega = -E_\alpha$. In particular, a sharp peak at $\omega=0$ directly signals the presence of the topological zero mode [top of Fig.\,\ref{fig:dynamics_spectrum}(a,b)].

In the trivial subspace, to probe the continuous spectrum supported by the one‑dimensional $|e\rangle|n\rangle$ chain, we initialize the system in the state $|\Psi_N(t=0)\rangle =|e,0\rangle$. In the weak‑coupling limit $r\ll g,s,\beta$, the subsequent time evolution remains predominantly confined to the $|e\rangle|n\rangle$ sites. Therefore, the spectral properties of the dynamics can be approximately analyzed using the chiral symmetry of this one-dimensional chain. The Hamiltonian of this chain is $\hat{H}_{e}=\sqrt{2}\beta |e\rangle\langle e|\hat{x}$, with eigenstates $|e, x\rangle$ and corresponding eigenvalues $E_{x}=\sqrt{2} \beta x$. Its chiral-symmetry operator is identified as
\begin{equation}
	\hat{C}_2 = (-1)^{\hat{a}^\dagger\hat{a}} \otimes |e\rangle\langle e|,
\end{equation}
satisfying $\hat{C}_{2}\hat{H}_{e}\hat{C}_{2}=-\hat{H}_{e}$ and $\hat{C}_{2}|e, x\rangle=|e, -x\rangle$. The initial state can be expanded in the eigenstates of $\hat{H}_{e}$ as
\begin{equation}
	|\Psi_N(t=0)\rangle=|e,0\rangle = \int_{-\infty}^{\infty} \psi(x) |e,x\rangle dx,
\end{equation}
where $\psi(x)=\langle x|n=0\rangle$ is the wave function of the Fock state $|0\rangle$ in the coordinate $\hat{x}$ representation. Since $\psi(x)$ is an even function, we have $\psi(-x) = \psi(x)$.The time-dependent expectation value of $\hat{C}_2$ then takes the form 
\begin{equation}
	\begin{split}
		&\langle \hat{C}_2(t) \rangle\\ &=\int_{-\infty}^{\infty}\int_{-\infty}^{\infty} \psi(x^\prime)^* \psi(x)\langle e, x^\prime\mid \hat{C}_2 \mid e,x\rangle e^{-i(E_x - E_{x'})t}dx dx'\\
		&=\int_{-\infty}^{\infty} |\psi(x)|^2 e^{-2i E_x t}dx.
	\end{split}
\end{equation}

 Performing a Fourier transform on $\langle \hat{C}_2(t) \rangle$ yields 
\begin{equation}
	\begin{split}
		F(\omega) &= \int_{-\infty}^{\infty} \langle \hat{C}_2(t) \rangle e^{-2i\omega t} dt \\
		&= \pi \int_{-\infty}^{\infty} |\psi(x)|^2 \delta(\omega + E_x) dx\\
		&=\frac{\pi}{\sqrt{2}|\beta|} \left| \psi\left(-\frac{\omega}{\sqrt{2}\beta} \right) \right|^2,
	\end{split}
\end{equation}
which corresponds to the continuous spectrum associated with the trivial subspace [middle of Fig.~\ref{fig:dynamics_spectrum}(a,b)].

To directly extract the BIC signature as shown in Fig.\,\ref{fig:BIC_FSL_SSH}(c), we prepare the system in the superposition state 
\begin{equation}
|\Psi(t=0)\rangle = \frac{1}{\sqrt{2}}\left(|\Psi_T(t=0)\rangle + |\Psi_N(t=0)\rangle\right),
\end{equation}
which has equal weight in both subspaces. Measuring the combined chiral operator $\hat{C}=\hat{C}_1+\hat{C}_2$ and Fourier-transforming its time evolution gives the full system spectrum, in which a sharp zero-mode peak is clearly visible embedded within the broad continuous background of the trivial subspace. This is the smoking-gun signature of the BIC [bottom of Fig.\,\ref{fig:dynamics_spectrum}(a,b)].

\section{experimental proposal with a single trapped ion}
\begin{figure}[t]
	\includegraphics[width=0.48\textwidth]{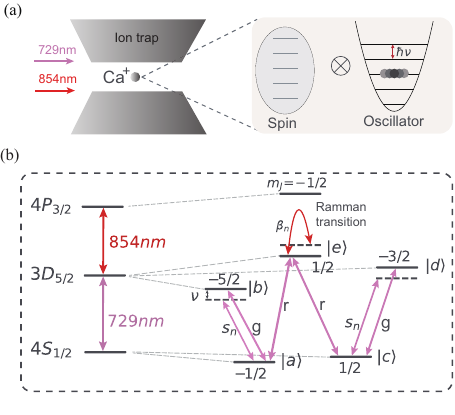}
	\centering
	\caption{\label{fig:experimental_setup}Experimental proposal for the quantum‑optical BIC with a trapped ion. (a) A single $^{40}\text{Ca}^+$ ion confined in a Paul trap, representing the composite spin‑boson system. The effective (pseudo‑)spin is encoded in a selected set of internal electronic states, and the bosonic mode corresponds to the harmonic motion (phonon) of the ion with frequency $\nu$. The system is coherently manipulated via 729‑\text{nm} and 854‑\text{nm} laser beams. (b) Relevant level scheme, optical transitions, and laser wavelengths for engineering the Hamiltonian that realizes the quantum‑optical BIC.}
\end{figure}

Here, we present an experimental proposal for realizing the quantum‑optical BIC using a single trapped ion. The scheme utilizes a single $^{40}\text{Ca}^+$ ion confined in a Paul trap [Figure \ref{fig:experimental_setup}(a)]. The internal degrees of freedom (multi‑level electronic states) are employed to encode a high‑dimensional pseudo‑spin, while the external degree of freedom (the harmonic motion of the ion’s center‑of‑mass) is quantized as a bosonic mode (phonon), thereby forming a composite spin‑boson system. The required spin‑level transitions and spin‑boson couplings are engineered via laser‑ion interactions at 729 \text{nm} and 854 \text{nm}.

Figure \ref{fig:experimental_setup}(b) illustrates the relevant level scheme, optical transitions, and laser wavelengths for implementing the Hamiltonian that yields the quantum‑optical BIC. We map the electronic states of the ion, $|4\mathrm{S}_{1/2}, m=-1/2\rangle$, $|3\mathrm{D}_{5/2}, m=-5/2\rangle$, $|4\mathrm{S}_{1/2}, m=1/2\rangle$, $|3\mathrm{D}_{5/2},$ $m=-3/2\rangle$, and $|3\mathrm{D}_{5/2}, m=1/2\rangle$, onto the model spin states $|a\rangle$, $|b\rangle$, $|c\rangle$, $|d\rangle$, and $|e\rangle$, respectively. The spin‑drive terms in the Hamiltonian are realized by resonantly driving the corresponding carrier transitions $|a\rangle \leftrightarrow |b\rangle$, $|a\rangle \leftrightarrow |e\rangle$, $|c\rangle \leftrightarrow |d\rangle$ and $|c\rangle \leftrightarrow |e\rangle$ with 729‑\text{nm} lasers, whose frequencies can be precisely tuned via acousto‑optic modulators (AOMs) to satisfy the resonance conditions. The two spin‑boson JC‑coupling terms are implemented by red‑detuning two lasers near 729\,\text{nm} by exactly one phonon frequency (i.\,e., satisfying the red‑sideband conditions) for the  transitions $|a\rangle \leftrightarrow |b\rangle$ and $|c\rangle \leftrightarrow |d\rangle$, respectively. Finally, the spin‑dependent bosonic drive term $\propto|e\rangle\langle e| \hat{x}$ is realized via a Raman process: two right‑circularly polarized 854‑\text{nm} lasers, which couple off‑resonantly to the $|4\mathrm{P}_{3/2}, m=-1/2\rangle$ level, are set with a frequency difference equal to the phonon frequency $\nu$, thereby resonantly driving the phonon mode only when the spin is in state $|e\rangle$.

The initial states $|\Psi_T(t=0)\rangle$, $|\Psi_N(t=0)\rangle$ and $|\Psi(t=0)\rangle$ are prepared by first cooling the phonon mode to its ground state via sideband cooling, followed by appropriate rotations on the high‑dimensional pseudo‑spin \cite{pseudospin}. The chiral‑symmetry operator for the spin degree $\hat{C}_1$ can be measured via sequential electron‑shelving fluorescence detection \cite{detection}, while the phonon chiral operator $\hat{C}_2$ can be probed using established phonon‑population reconstruction techniques \cite{phononcons}. The expectation value of the composite operator $\hat{C}$ is obtained by measuring the two parts independently and summing the results. With typical JC-coupling strength ($g\sim2\pi \times 10 \text{KHz}$), the total evolution time $g t=200 $ in Fig.\,\ref{fig:dynamics_spectrum} remains below 5 \text{ms}, which is shorter than the typical spin and phonon coherence times achieved in current ion‑trap experiments \cite{experi}. Hence, the proposed scheme is highly feasible on existing trapped‑ion platforms.

\section{conclusion}In conclusion, we introduced the concept of a quantum-optical BIC and constructed a paradigmatic model to realize it within quantum optics. This model requires only a few physical degrees of freedom, yet it clearly reveals the quantum interference mechanism underlying BICs from first principles. Experimentally, the model can be implemented using a synthetic FSL engineered with a single trapped ion. Our work opens a new avenue that bridges the actively studied field of BICs with quantum optics—a discipline endowed with mature experimental techniques—thus offering a wealth of promising platforms for investigating BICs. Given the close connection between quantum-optical models and quantum information science as well as quantum metrology, a natural and compelling direction for future research is to explore whether the exotic properties and rich applications of classical optical BICs can be transferred to this quantum-optical framework, thereby substantially advancing the development of quantum BICs.

\begin{acknowledgments}
This work was supported by the National Natural Science Foundation of China (Grants No. 12174447, No. 12174448, and No. 11574398), Quantum Science and Technology-National Science and Technology Major Project (QNMP) (Grant No.2021ZD0301605), the Science and Technology Innovation Program of Hunan under Grant No. 2022RC1194, and the Natural Science Foundation of Hunan Province under Grant No. 2023JJ10052.

\end{acknowledgments}

\bibliography{ref}% Produces the bibliography via BibTeX.

\end{document}